\documentclass[aps,prl,twocolumn,superscriptaddress,showpacs]{revtex4} 
\usepackage{graphicx}
\usepackage{dcolumn}
\usepackage{bm}
\usepackage{setspace}
\def\ppbar{$p\overline{p}~$}             

\def\met{\mbox{${\hbox{$E$\kern-0.6em\lower-.1ex\hbox{/}}}_T$}} 
\def\mex{\mbox{${\hbox{$E$\kern-0.6em\lower-.1ex\hbox{/}}}_x$}} 
\def\mey{\mbox{${\hbox{$E$\kern-0.6em\lower-.1ex\hbox{/}}}_y$}} 
\def\mexy{\mbox{${\hbox{$E$\kern-0.6em\lower-.1ex\hbox{/}}}_{x,y}$}} 
 
\def\D0{D\O}                            
\def\etal{{\sl et al.}}                 
\newcommand{\pythia}{{\sc pythia}}

\newcommand{\geant}{{\sc geant}}

\begin{document}

\title{
\Large\bf
Measurement of the Inclusive Isolated Prompt Photon Cross \\ 
Section in \ppbar\ Collisions at $\sqrt{s} = 1.96$~TeV using the CDF Detector
}

\affiliation{Institute of Physics, Academia Sinica, Taipei, Taiwan 11529, Republic of China} 
\affiliation{Argonne National Laboratory, Argonne, Illinois 60439} 
\affiliation{University of Athens, 157 71 Athens, Greece} 
\affiliation{Institut de Fisica d'Altes Energies, Universitat Autonoma de Barcelona, E-08193, Bellaterra (Barcelona), Spain} 
\affiliation{Baylor University, Waco, Texas  76798} 
\affiliation{Istituto Nazionale di Fisica Nucleare Bologna, $^{cc}$University of Bologna, I-40127 Bologna, Italy} 
\affiliation{Brandeis University, Waltham, Massachusetts 02254} 
\affiliation{University of California, Davis, Davis, California  95616} 
\affiliation{University of California, Los Angeles, Los Angeles, California  90024} 
\affiliation{University of California, San Diego, La Jolla, California  92093} 
\affiliation{University of California, Santa Barbara, Santa Barbara, California 93106} 
\affiliation{Instituto de Fisica de Cantabria, CSIC-University of Cantabria, 39005 Santander, Spain} 
\affiliation{Carnegie Mellon University, Pittsburgh, PA  15213} 
\affiliation{Enrico Fermi Institute, University of Chicago, Chicago, Illinois 60637}
\affiliation{Comenius University, 842 48 Bratislava, Slovakia; Institute of Experimental Physics, 040 01 Kosice, Slovakia} 
\affiliation{Joint Institute for Nuclear Research, RU-141980 Dubna, Russia} 
\affiliation{Duke University, Durham, North Carolina  27708} 
\affiliation{Fermi National Accelerator Laboratory, Batavia, Illinois 60510} 
\affiliation{University of Florida, Gainesville, Florida  32611} 
\affiliation{Laboratori Nazionali di Frascati, Istituto Nazionale di Fisica Nucleare, I-00044 Frascati, Italy} 
\affiliation{University of Geneva, CH-1211 Geneva 4, Switzerland} 
\affiliation{Glasgow University, Glasgow G12 8QQ, United Kingdom} 
\affiliation{Harvard University, Cambridge, Massachusetts 02138} 
\affiliation{Division of High Energy Physics, Department of Physics, University of Helsinki and Helsinki Institute of Physics, FIN-00014, Helsinki, Finland} 
\affiliation{University of Illinois, Urbana, Illinois 61801} 
\affiliation{The Johns Hopkins University, Baltimore, Maryland 21218} 
\affiliation{Institut f\"{u}r Experimentelle Kernphysik, Karlsruhe Institute of Technology, D-76131 Karlsruhe, Germany} 
\affiliation{Center for High Energy Physics: Kyungpook National University, Daegu 702-701, Korea; Seoul National University, Seoul 151-742, Korea; Sungkyunkwan University, Suwon 440-746, Korea; Korea Institute of Science and Technology Information, Daejeon 305-806, Korea; Chonnam National University, Gwangju 500-757, Korea; Chonbuk National University, Jeonju 561-756, Korea} 
\affiliation{Ernest Orlando Lawrence Berkeley National Laboratory, Berkeley, California 94720} 
\affiliation{University of Liverpool, Liverpool L69 7ZE, United Kingdom} 
\affiliation{University College London, London WC1E 6BT, United Kingdom} 
\affiliation{Centro de Investigaciones Energeticas Medioambientales y Tecnologicas, E-28040 Madrid, Spain} 
\affiliation{Massachusetts Institute of Technology, Cambridge, Massachusetts  02139} 
\affiliation{Institute of Particle Physics: McGill University, Montr\'{e}al, Qu\'{e}bec, Canada H3A~2T8; Simon Fraser University, Burnaby, British Columbia, Canada V5A~1S6; University of Toronto, Toronto, Ontario, Canada M5S~1A7; and TRIUMF, Vancouver, British Columbia, Canada V6T~2A3} 
\affiliation{University of Michigan, Ann Arbor, Michigan 48109} 
\affiliation{Michigan State University, East Lansing, Michigan  48824}
\affiliation{Institution for Theoretical and Experimental Physics, ITEP, Moscow 117259, Russia} 
\affiliation{University of New Mexico, Albuquerque, New Mexico 87131} 
\affiliation{Northwestern University, Evanston, Illinois  60208} 
\affiliation{The Ohio State University, Columbus, Ohio  43210} 
\affiliation{Okayama University, Okayama 700-8530, Japan} 
\affiliation{Osaka City University, Osaka 588, Japan} 
\affiliation{University of Oxford, Oxford OX1 3RH, United Kingdom} 
\affiliation{Istituto Nazionale di Fisica Nucleare, Sezione di Padova-Trento, $^{dd}$University of Padova, I-35131 Padova, Italy} 
\affiliation{LPNHE, Universite Pierre et Marie Curie/IN2P3-CNRS, UMR7585, Paris, F-75252 France} 
\affiliation{University of Pennsylvania, Philadelphia, Pennsylvania 19104}
\affiliation{Istituto Nazionale di Fisica Nucleare Pisa, $^{ee}$University of Pisa, $^{ff}$University of Siena and $^{gg}$Scuola Normale Superiore, I-56127 Pisa, Italy} 
\affiliation{University of Pittsburgh, Pittsburgh, Pennsylvania 15260} 
\affiliation{Purdue University, West Lafayette, Indiana 47907} 
\affiliation{University of Rochester, Rochester, New York 14627} 
\affiliation{The Rockefeller University, New York, New York 10021} 
\affiliation{Istituto Nazionale di Fisica Nucleare, Sezione di Roma 1, $^{hh}$Sapienza Universit\`{a} di Roma, I-00185 Roma, Italy} 

\affiliation{Rutgers University, Piscataway, New Jersey 08855} 
\affiliation{Texas A\&M University, College Station, Texas 77843} 
\affiliation{Istituto Nazionale di Fisica Nucleare Trieste/Udine, I-34100 Trieste, $^{ii}$University of Trieste/Udine, I-33100 Udine, Italy} 
\affiliation{University of Tsukuba, Tsukuba, Ibaraki 305, Japan} 
\affiliation{Tufts University, Medford, Massachusetts 02155} 
\affiliation{Waseda University, Tokyo 169, Japan} 
\affiliation{Wayne State University, Detroit, Michigan  48201} 
\affiliation{University of Wisconsin, Madison, Wisconsin 53706} 
\affiliation{Yale University, New Haven, Connecticut 06520} 
\author{T.~Aaltonen}
\affiliation{Division of High Energy Physics, Department of Physics, University of Helsinki and Helsinki Institute of Physics, FIN-00014, Helsinki, Finland}
\author{J.~Adelman}
\affiliation{Enrico Fermi Institute, University of Chicago, Chicago, Illinois 60637}
\author{B.~\'{A}lvarez~Gonz\'{a}lez$^v$}
\affiliation{Instituto de Fisica de Cantabria, CSIC-University of Cantabria, 39005 Santander, Spain}
\author{S.~Amerio$^{dd}$}
\affiliation{Istituto Nazionale di Fisica Nucleare, Sezione di Padova-Trento, $^{dd}$University of Padova, I-35131 Padova, Italy} 

\author{D.~Amidei}
\affiliation{University of Michigan, Ann Arbor, Michigan 48109}
\author{A.~Anastassov}
\affiliation{Northwestern University, Evanston, Illinois  60208}
\author{A.~Annovi}
\affiliation{Laboratori Nazionali di Frascati, Istituto Nazionale di Fisica Nucleare, I-00044 Frascati, Italy}
\author{J.~Antos}
\affiliation{Comenius University, 842 48 Bratislava, Slovakia; Institute of Experimental Physics, 040 01 Kosice, Slovakia}
\author{G.~Apollinari}
\affiliation{Fermi National Accelerator Laboratory, Batavia, Illinois 60510}
\author{A.~Apresyan}
\affiliation{Purdue University, West Lafayette, Indiana 47907}
% ...

\collaboration{CDF Collaboration\footnote{With visitors from $^a$University of Massachusetts Amherst, Amherst, Massachusetts 01003,
$^b$Universiteit Antwerpen, B-2610 Antwerp, Belgium, 
$^c$University of Bristol, Bristol BS8 1TL, United Kingdom,
$^d$Chinese Academy of Sciences, Beijing 100864, China, 
$^e$Istituto Nazionale di Fisica Nucleare, Sezione di Cagliari, 09042 Monserrato (Cagliari), Italy,
$^f$University of California Irvine, Irvine, CA  92697, 
$^g$University of California Santa Cruz, Santa Cruz, CA  95064, 
$^h$Cornell University, Ithaca, NY  14853, 
$^i$University of Cyprus, Nicosia CY-1678, Cyprus, 
$^j$University College Dublin, Dublin 4, Ireland,
$^k$University of Edinburgh, Edinburgh EH9 3JZ, United Kingdom, 
$^l$University of Fukui, Fukui City, Fukui Prefecture, Japan 910-0017
$^m$Kinki University, Higashi-Osaka City, Japan 577-8502
$^n$Universidad Iberoamericana, Mexico D.F., Mexico,
$^o$University of Iowa, Iowa City, IA  52242,
$^p$Kansas State University, Manhattan, KS 66506
$^q$Queen Mary, University of London, London, E1 4NS, England,
$^r$University of Manchester, Manchester M13 9PL, England,
$^s$Muons, Inc., Batavia, IL 60510, 
$^t$Nagasaki Institute of Applied Science, Nagasaki, Japan, 
$^u$University of Notre Dame, Notre Dame, IN 46556,
$^v$University de Oviedo, E-33007 Oviedo, Spain, 
$^w$Texas Tech University, Lubbock, TX  79609, 
$^x$IFIC(CSIC-Universitat de Valencia), 56071 Valencia, Spain,
$^y$Universidad Tecnica Federico Santa Maria, 110v Valparaiso, Chile,
$^z$University of Virginia, Charlottesville, VA  22906
$^{aa}$Bergische Universit\"at Wuppertal, 42097 Wuppertal, Germany,
$^{bb}$Yarmouk University, Irbid 211-63, Jordan
$^{jj}$On leave from J.~Stefan Institute, Ljubljana, Slovenia, 
}}
\noaffiliation


\date{
September 9, 2009
}

\begin{abstract}
A measurement of the cross section for the inclusive production of
isolated photons by the CDF experiment at the Fermilab Tevatron collider
is presented.  The measurement covers the pseudorapidity region
$|\eta^{\gamma}|<1.0$ and the transverse energy range $E_T^\gamma>30$
GeV and is based on 2.5 fb$^{-1}$ of integrated luminosity.
The sample is almost a factor of seven larger than those used for recent 
published results and extends the $E_T^{\gamma}$ coverage by 100 GeV.
The result agrees with next-to-leading order perturbative QCD 
calculations within uncertainties over the range $50<E_T^\gamma<400$ GeV, 
though the energy spectrum in the data shows a steeper slope at lower $E_T^\gamma$.
\end{abstract}

\maketitle


The measurement of inclusive prompt-photon production constitutes a
test of perturbative QCD (pQCD) with the potential to constrain parton
distribution functions (PDF), while avoiding the complications of jet
identification and energy measurements.  The photon cross section is
also sensitive to the presence of new physics at large photon transverse
energy. In high-energy \ppbar\ collisions, photons are mostly produced
via quark--gluon Compton scattering or quark--anti-quark annihilation.
In addition to these processes, photons can also be produced through the
fragmentation of outgoing partons, though this contribution is reduced when
the photon is required to be isolated from other particles in the final
state. An isolation requirement is necessary to suppress the background
from energetic $\pi^0$ and $\eta$ mesons.

Previous experiments have presented results on inclusive prompt
photon production compared to next-to-leading order (NLO)
pQCD predictions~\cite{ref:cdfrun1, ref:d0run1, ref:other, ref:d0run2}.
In this Letter, we present a new measurement of the inclusive cross section
based on 2.5 fb$^{-1}$ of integrated luminosity collected by the CDF~II
detector at the Tevatron.  Photons are required to be isolated, have a
pseudorapidity $|\eta^{\gamma}|<1.0$, and a transverse energy $E_T^{\gamma}>
30$ GeV~\cite{ref:coord}.  This measurement extends the $E_T^{\gamma}$ coverage by
$100$ GeV compared to previous measurements~\cite{ref:d0run2}, and presents
reduced systematic uncertainties as a result of using an improved background
subtraction method based on the shape of the isolation distribution.

The CDF II detector~\cite{ref:CDF_detector} is a general-purpose particle
detector located at the Tevatron collider.
The charged-particle tracking system is immersed in a $1.4$ T
magnetic field aligned coaxially with the beam line, and provides tracking
coverage in the pseudorapidity range $|\eta|\le 2.0$.
A central preradiator detector surrounds the tracking system
and samples the electromagnetic showers that begin in the material
in front of it. This detector consisted of multi-wire proportional
chambers at the beginning of Run~II, and was upgraded to scintillation
tiles in 2004.
Scintillator-based electromagnetic (EM)
and hadronic (HAD) calorimeters arranged in projective towers of
size $\Delta\eta \times \Delta\phi=0.1 \times 0.26$
provide a 
coverage of $|\eta| < 3.6$.
The energy resolution of the CEM calorimeters for photons and electrons is 
$\sigma/E_T=13.5\%/\sqrt{E_T}\oplus 2\%$, where $\oplus$ represents
sum in quadrature.
The central ($|\eta| < 1.1$) electromagnetic strip chambers (CES)
are multi-wire proportional chambers embedded inside the EM calorimeter
and positioned at a depth corresponding to the expected maximum of the longitudinal
shower profile (6 radiation lengths). Anode wires
and cathode strips measure $\phi$ and $z$ respectively, providing
a $2$ mm position resolution in each direction for 50 GeV electrons.


The data are collected using a 
three-level on-line event selection system (trigger)
that selects events
with at least one energy cluster consistent with a photon in the final state.
A photon cluster consists of one to three consecutive calorimeter towers in the 
$\eta$ direction.
Photons are collected with two trigger thresholds
in $E_T^{\gamma}$, 25 GeV and 70 GeV. In order to reduce
contamination from neutral meson decays, the low $E_T^{\gamma}$
trigger requires photon clusters to be isolated.
The isolation requirement at the second level uses simple box patterns of
towers, while at the third level
the extra energy inside
a cone of radius $R=\sqrt{(\Delta\phi)^2+(\Delta\eta)^2}=0.4$ around the
cluster is required to be less than $10\%$ of the energy of the cluster
($E^{clu}$).
The low $E_T^{\gamma}$
trigger also requires the lateral shower profile of the CES cluster to be
consistent with that of electrons, as measured in test beam data.

The event selection requires the primary vertex $z$ position to be
within $60$ cm of the center of the detector to maintain the projective
nature of the calorimeter towers.
In order to suppress beam-related backgrounds, cosmic rays,
and calorimeter noise, as well as leptonic
$W$ decays~\cite{ref:metcut},
the missing transverse energy of the event has to be less than 
$80\%$ of the transverse energy of the leading photon candidate.
This requirement reduces
these backgrounds to less than one percent each,
while preserving almost $99\%$ of the photon signal.
Photon candidates are required to be matched to
a photon cluster and to be fiducial to the CES detector.
Photons with additional CES clusters are rejected, since
contributions from neutral mesons occasionally produce multiple clusters.
At most, one low transverse momentum track ($p_T^{trk}<1$GeV$/c+0.5\% E_T^{\gamma}/c$)
is allowed to point to the photon candidate.
The fraction of the energy of the photon in the hadronic
calorimeter has to be small ($E^{HAD}/E^{EM}<0.055+0.00045 E^{clu}/\rm GeV$),
and, for events collected using the low $E_T$ trigger, the CES shower associated with the photon
candidate has to have a profile consistent with test beam electrons~\cite{ref:ces}.

The transverse energy of the photon is
corrected to
account for non-uniformities in the calorimeter response, and
calibrated using electrons from reconstructed $Z$ bosons.
Photon candidates
are required to have $E_T^\gamma>30$ GeV and to be isolated in the
calorimeter, $E_T^{R=0.4}-E_T^\gamma=E_T^{iso}<2$ GeV,
where $E_T^{R=0.4}$ is the transverse energy in a cone
of $R=0.4$ around the photon.
The isolation requirement reduces the background from neutral
mesons, but also suppresses the photon signal coming from
parton fragmentation processes.

\begin{figure}[htb]
\begin{center}
\includegraphics[width=3.2in]{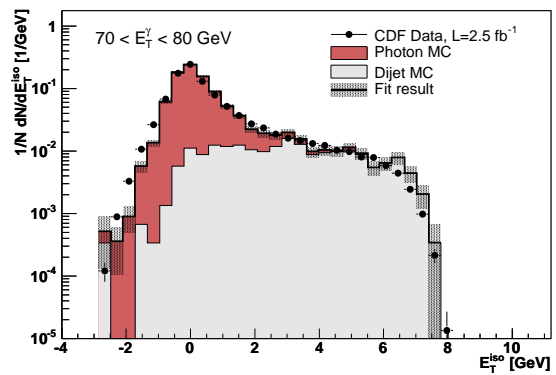}
\caption{Measured isolation distribution for 
photons with $70<E_T^\gamma<80$ GeV. A $\chi^2$
fit to the data (full points) with photon (dark histogram) 
and neutral mesons (light histogram)
templates is used to extract the
photon fraction. The result of the fit is shown as
a full line with the associated uncertainty from the fitting procedure.
}
\label{fig:fit}
\end{center}
\end{figure}

\begin{center}
\begin{figure}[htb]
\includegraphics[width=3.2in]{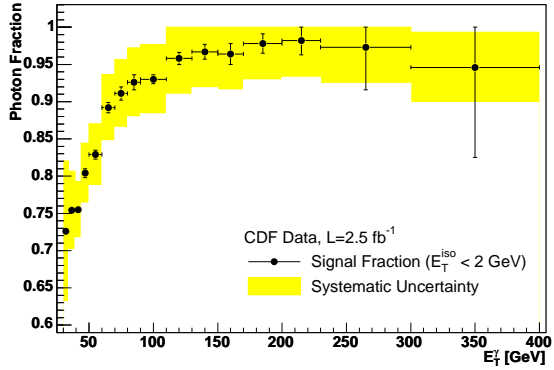}
\caption{Fraction of isolated prompt
photons as a function of  $E_T^{\gamma}$.
The systematic uncertainty band is discussed
in the text.
}
\label{fig:sf}
\end{figure}
\end{center}

While the selection criteria remove the bulk of the
neutral meson background, substantial contamination remains,
mainly corresponding to fluctuations in the fragmentation of jets, leading to
neutral mesons that carry most of the parton energy.
To subtract these isolated $\pi^0$
and $\eta$ mesons, the isolation distribution from the data (without
the $E_T^{iso}<2$ GeV requirement) is fitted with signal and background
templates formed using Monte Carlo simulation.
The isolation distribution is sensitive to the differences
between prompt photons and background: photons produce a
well-defined peak at low isolation while neutral mesons present a
flatter shape.
The signal template is obtained from a
{\sc pythia} 6.216~\cite{ref:pythia} photon Monte Carlo sample and the
background template is constructed by selecting photons from meson decays in a QCD
\pythia\ sample.
In both cases the underlying event model of {\sc pythia}
is tuned to CDF jet data ({\sc tune a}~\cite{ref:tunes}).
These events are passed through a \geant\cite{ref:geant} simulation of
the detector and subjected to the same selection requirements as the data.
As an example, Fig.~\ref{fig:fit} shows the isolation distribution in data
compared to signal and background templates for photons in the region
$70 < E_T^\gamma < 80$ GeV.  
The raw $E_T^{iso}$ is corrected for leakage effects
and pile-up contributions which occasionally yields a negative
$E_T^{iso}$, but if the negative $E_T^{iso}$ bins 
are not included in the fit, the photon fraction changes by less than $2\%$.
At high $E_T^\gamma$ ($>200$ GeV) the signal template,
as extracted from the Monte Carlo simulation, does not describe accurately
the signal peak. This is attributed to deficiencies in the details of the
shower simulation in the calorimeter and, to a lesser extent, to the model
of the underlying event~\cite{ref:thesis_carolina}.
An $E_T^\gamma$-dependent correction is applied to modify the signal templates
and improve the fitting process. At low $E_T^\gamma$ no correction is necessary
while at very high $E_T^\gamma$ the template is shifted by $-0.5$~GeV and its
width reduced by $50\%$. The correction, which is not applied to the background 
templates, only changes the final results by few percent, and is accounted
for in the study of systematic uncertainties (see below).
As a result of the fitting procedure, the photon fraction (${\cal F}$)
is extracted from the measured isolation distributions. Figure~\ref{fig:sf}
shows the values for ${\cal F}$ as a function of $E_T^\gamma$,
including the systematic uncertainties discussed below.

The raw inclusive differential cross section as a function of $E_{T}^\gamma$ is defined as
$d\sigma/(dE_{T}^\gamma\,d\eta^\gamma)=(N^\gamma\,{\cal F})/(\Delta E_{T}^\gamma\,\Delta\eta^\gamma\,\epsilon_{trig} \,{\cal L})$,
where $N^\gamma$ is the number of photon candidates in a given $E_{T}^\gamma$ bin,
$\Delta E_{T}^\gamma$ ($\Delta\eta^\gamma$) is the size of the $E_{T}^\gamma$ ($\eta^\gamma$) bin,
$\epsilon_{trig}$ the trigger efficiency, and ${\cal L}$ is the integrated
luminosity.
The trigger efficiency is approximately $100\%$ in the kinematic region of the measurement.
The measured cross section is corrected for acceptance, efficiency of the photon selection,
and resolution effects back to the hadron level~\cite{ref:hadron} using a bin-by-bin unfolding
procedure and a sample of prompt-photon events simulated with {\sc pythia}.
To avoid any bias on the unfolding factors due to
assumptions about the true  $E_{T}^\gamma$ spectrum,
the photon Monte Carlo sample is re-weighted
to match the measured spectrum. The resulting unfolding
factors vary between $0.638\pm0.003$ and $0.69\pm0.01$ with little $E_{T}^\gamma$
dependence.

A detailed study of systematic uncertainties is carried out~\cite{ref:thesis_carolina}.
The largest contribution to the total
uncertainty at high $E_T^\gamma$ is caused by the $1.5\%$ uncertainty
on the photon absolute energy scale,
due to a small energy dependence in the energy ratio
of simulated and data electrons from the $Z$ mass peak. 
This introduces an uncertainty on the measured cross section
that varies between $6\%$ and $13\%$ as $E_T^\gamma$ increases.
At low $E_T^\gamma$ the dominant uncertainty source is the photon fraction.
Different methods are considered to construct signal and background templates
and extract ${\cal F}$: signal templates are defined using electrons
from $Z$ decays in data instead of using Monte Carlo simulated events;
very simple templates (two bins in isolation) are considered, to remove
the details of the isolation distribution in the fitting
procedure; the photon signal is then extracted using  
background templates with and without the $E_{T}^\gamma$-dependent correction 
that is applied to the signal templates.
In addition,  a completely different method~\cite{ref:ces} based on
the shower profile of the photon candidate in the CES detector and the
number of conversions in the material in front of the preradiator
detector is used to determine the neutral meson background.  As a result,
a conservative systematic uncertainty that varies between $13\%$ at low
$E_T^\gamma$ and $5\%$ at high $E_T^\gamma$ (see Fig.~\ref{fig:sf}) is
assigned to the measured photon signal, covering the results from all
the alternative methods. A $3\%$ uncertainty on the measured cross section,
approximately independent of $E_T^\gamma$, reflects uncertainties on the
determination of the photon acceptance, and a  $5\%$ uncertainty on the
measured cross section at low $E_T^\gamma$ accounts for the uncertainty
on the CES cut efficiency.
An additional $10\%$ uncertainty in the photon isolation energy 
introduces a $1\%$ uncertainty on the measurement.
Performing the unfolding procedure using unweighted Monte Carlo samples
resulted in a less than $1\%$ effect on the measured cross section.
The different sources of systematic uncertainty are added in quadrature. The
total systematic uncertainty varies between $15\%$ at low and very high
$E_T^\gamma$, and $8\%$ at intermediate $E_T^\gamma$.  Finally, an additional
$6\%$ uncertainty due to the measurement of the integrated luminosity is 
considered~\cite{ref:lum}.

\begin{center}
\begin{figure}[htb]
\includegraphics[width=3.2in]{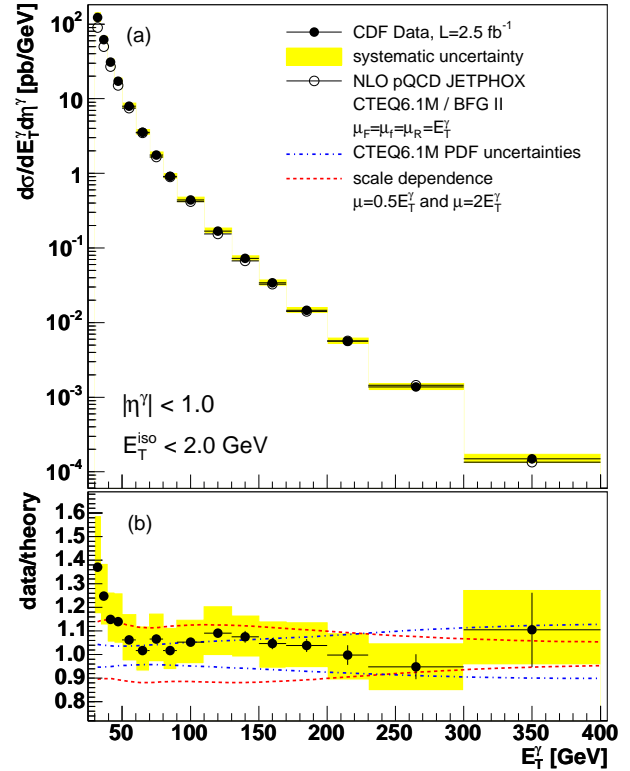}
\caption{(a) Measured inclusive isolated prompt-photon cross section as a function of $E_T^\gamma$ 
compared to NLO pQCD predictions. (b) Ratio data/theory as a function of
$E_T^\gamma$. The shaded band includes the total systematic uncertainty
on the measurement except for the $6\%$ luminosity uncertainty. 
The dashed and dotted lines indicate the PDF
uncertainty and the variation with NLO pQCD predictions,
respectively. A parton-to-hadron correction, $C_{had}=0.91\pm0.03$, is applied to the theoretical
predictions.
}
\label{fig:xsec}
\end{figure}
\end{center}

The measured inclusive isolated prompt photon cross section as
a function of $E_T^\gamma$ is presented in
Fig.~\ref{fig:xsec} and Table~\ref{tab:result}.
The data are compared to NLO pQCD predictions as determined
by the {\sc jetphox}~\cite{ref:jetphox} program with CTEQ6.1M
PDF~\cite{ref:cteq6m}, normalization, factorization
and fragmentation scales set to $E_T^\gamma$, and photon
isolation requirement as for the data.
Variations of the scales by a factor of two change the prediction by $15\%$
at low $E_T^\gamma$ and $8\%$ at high $E_T^\gamma$.  The uncertainty on the predictions
due to PDF varies between $4\%$ at low $E_T^\gamma$ and $13\%$ at
high $E_T^\gamma$, as determined using the Hessian method~\cite{ref:hessian}.
In addition, we have evaluated the theoretical prediction using the
MRST04~\cite{ref:mrst04} PDF, and find it well inside the experimental and
other theoretical uncertainties.

The theoretical prediction includes an additional correction factor,
$C_{had}(E_T^{\gamma})$, to account for the presence of non-pQCD contributions 
from the underlying event and fragmentation into hadrons,
that tend to increase the energy in the isolation cone.
$C_{had}$ is estimated, using Monte Carlo generated events,  as the ratio
between the nominal $E_T^{\gamma}$ distribution at the hadron level and
the one obtained after turning off both the interactions between proton
and antiproton remnants and the string fragmentation in the Monte Carlo
samples.  Two different sets of  tuned parameters in \pythia\ ({\sc tune
a} and {\sc dw}~\cite{ref:tunes}) are considered,  and the mean effect
$C_{had}=0.91\pm0.03$ , observed to have little $E_T^{\gamma}$ dependence,  is taken
as the correction. The uncertainty on $C_{had}$ covers the results obtained
with the different \pythia\ tunes.  As expected, the correction reduces the
predicted cross section, since the presence of underlying event activity
results in photons failing the isolation requirement.

A difference between data and the NLO pQCD predictions is observed for
$E_T^\gamma<50$~GeV.
Discrepancies were also observed at low $p_T$ in previous measurements
at collider and fixed target experiments~\cite{ref:cdfrun1, ref:d0run1,
ref:other}.  For $E_T^\gamma>50$~GeV, good agreement is observed, and a global
$\chi^2$ test in this region, including correlations between systematic
uncertainties across $E_T^\gamma$ bins, finds a probability of 21\%. 

In conclusion, the measurement of the cross section for the inclusive
production of isolated prompt photons with $30<E_T^\gamma<400$~GeV and
$|\eta^{\gamma}|<1.0$ in \ppbar\ collisions at $\sqrt{s}=1.96$ TeV, 
using data corresponding to an integrated luminosity of
2.5~fb$^{-1}$, has been presented.  The sample is almost a factor of
seven larger than those used for recent published results~\cite{ref:d0run2} and extends
the $E_T^{\gamma}$ coverage by 100 GeV.  A new method to determine the
photon fraction based on the shape of the isolation distribution was
implemented for the first time at CDF, which resulted in smaller
uncertainties compared to previous results~\cite{ref:cdfrun1}.  The measured
cross section agrees with NLO pQCD predictions within uncertainties except
at low  $E_T^\gamma$ where the data have a steeper slope.

We thank the Fermilab staff and the technical staffs of the participating
institutions for their vital contributions. This work was supported by
the U.S. Department of Energy and National Science Foundation; the Italian
Istituto Nazionale di Fisica Nucleare; the Ministry of Education, Culture,
Sports, Science and Technology of Japan; the Natural Sciences and Engineering
Research Council of Canada; the National Science Council of the Republic of
China; the Swiss National Science Foundation; the A.P. Sloan Foundation;
the Bundesministerium f\"ur Bildung und Forschung, Germany; the World
Class University Program, the National Research Foundation of Korea; the
Science and Technology Facilities Council and the Royal Society, UK; the
Institut National de Physique Nucleaire et Physique des Particules/CNRS;
the Russian Foundation for Basic Research; the Ministerio de Ciencia e
Innovaci\'{o}n, and Programa Consolider-Ingenio 2010, Spain; the Slovak
R\&D Agency; and the Academy of Finland.

%
\begingroup
\begin{table}[hbt]
\footnotesize
\begin{tabular}{cc@{$\times$}lr@{,}r}\hline
{$E_T^{\gamma}$} & \multicolumn{2}{c}{$d\sigma/dE_T^{\gamma} d\eta^{\gamma}$} & \multicolumn{2}{c}{Syst. Unc.} \\
(GeV)                 & \multicolumn{2}{c}{(pb/GeV)}                               & \multicolumn{2}{c}{($\%$)}
\\ \hline
30--34   & (1.23$\pm$0.01)&10$^2   $ & ${+15.5}$&${-14.5}$  \\
34--39   & (6.21$\pm$0.03)&10$^1   $ & ${+10.8}$&${ -9.8}$  \\
39--44   & (3.10$\pm$0.02)&10$^1   $ & ${ +9.8}$&${ -8.4}$  \\
44--50   & (1.72$\pm$0.02)&10$^1   $ & ${+10.2}$&${ -8.1}$  \\
50--60   & (7.93$\pm$0.08)&10$^0   $ & ${+10.1}$&${ -8.4}$  \\
60--70   & (3.54$\pm$0.05)&10$^0   $ & ${ +9.8}$&${ -8.5}$  \\
70--80   & (1.76$\pm$0.03)&10$^0   $ & ${+10.0}$&${ -9.1}$  \\
80--90   & (9.08$\pm$0.14)&10$^{-1}$ & ${ +9.3}$&${ -7.9}$  \\
90--110  & (4.41$\pm$0.05)&10$^{-1}$ & ${ +8.8}$&${ -8.7}$  \\
110--130 & (1.68$\pm$0.03)&10$^{-1}$ & ${ +8.6}$&${ -8.7}$  \\
130--150 & (7.25$\pm$0.16)&10$^{-2}$ & ${ +7.8}$&${ -8.0}$  \\
150--170 & (3.41$\pm$0.08)&10$^{-2}$ & ${ +8.8}$&${-10.0}$  \\
170--200 & (1.46$\pm$0.04)&10$^{-2}$ & ${ +8.8}$&${ -9.1}$  \\
200--230 & (5.66$\pm$0.24)&10$^{-3}$ & ${ +9.0}$&${-10.6}$  \\
230--300 & (1.38$\pm$0.08)&10$^{-3}$ & ${+10.0}$&${-10.7}$  \\
300--400 & (1.49$\pm$0.21)&10$^{-4}$ & ${+15.2}$&${-13.4}$  \\
\hline
\end{tabular}
\caption{
\label{tab:result}
\small Measured inclusive isolated prompt photon cross section for photons in the
pseudorapidity region $|\eta^{\gamma}|<1.0$ and $30<E_T^\gamma<400$~GeV. 
The uncertainties in the central column are statistical. 
The additional $6\%$ luminosity uncertainty is not included in the table.
A parton-to-hadron correction ($C_{had}=0.91\pm0.03$) is applied to the pQCD
predictions.
}
\end{table}
\endgroup

\begin{thebibliography}{99}
\bibitem{ref:coord}
A cylindrical coordinate system with the $z$ axis along the proton
direction is used, in which $\theta$ is the polar angle. We define
$E_T=E\,\sin\theta$, $p_T=p\,\sin\theta$, and pseudorapidity 
$\eta= -\ln(\tan(\theta/2))$. 
%
The missing transverse energy is defined by 
$\met=-\sum_i E_T^i \hat{n}_i$, where $i$ is the calorimeter tower number and 
$\hat{n}_i$  is a unit vector perpendicular to the beam axis and pointing
at the $i^{th}$ calorimeter tower. The missing transverse energy 
as measured by the calorimeters 
is corrected for jets and muons.
\bibitem{ref:metcut}
Cosmic rays may produce  bremsstrahlung
photons. Muons from beam-halo interactions with the beam pipe
may in turn interact with the detector material
producing photons. Electrons from $W$ decays may fake photons
if the associated track is not reconstructed. All these contributions
lead to large missing transverse energy.
\bibitem{ref:hadron} The hadron level in the Monte Carlo generators
is defined using all final-state particles with lifetimes above $10^{-11}$~s.
\end{thebibliography}
\end{document}